\documentclass[iop]{emulateapj}

\usepackage{graphicx}
\usepackage{graphics}
\usepackage{txfonts}
\usepackage{epsfig}
\usepackage{natbib}
\usepackage{times}
\usepackage{amssymb}
\usepackage{color}
\usepackage{longtable}

\newcommand\ergcms{erg\,cm$^{-2}$\,s$^{-1}$}

\newcommand\ergs{erg\,s$^{-1}$}

\newcommand\cmsq{cm$^{-2}$}

\newcommand\chan{{\it{Chandra}}}

\newcommand\ymc{VVV CL077}
\newcommand\nh{$N_\mathrm{H}$}

\shorttitle{The X-ray population of VVV~CL077}
\shortauthors{Bodaghee et al.}

\begin{document}

\title{A first look at the X-ray population of the young massive cluster VVV~CL077}

\author{Arash Bodaghee\altaffilmark{1,2}}
\author{John A. Tomsick\altaffilmark{1}}
\author{Francesca Fornasini\altaffilmark{1}}
\author{Farid Rahoui\altaffilmark{3,4}}
\author{Franz E. Bauer\altaffilmark{5,6,7}}

\altaffiltext{1}{Space Sciences Laboratory, 7 Gauss Way, University of California, Berkeley, CA 94720, USA}
\altaffiltext{2}{Dept. of Chemistry, Physics, and Astronomy, Georgia College and State University, CBX 082, Milledgeville, GA 31061, USA}
\altaffiltext{3}{European Southern Observatory, Karl-Schwarzschild-Strasse 2, 85748 Garching bei M\"{u}nchen, Germany}
\altaffiltext{4}{Dept. of Astronomy, Harvard University, 60 Garden Street, Cambridge, MA 02138, USA}
\altaffiltext{5}{Instituto de Astrof\'{i}sica, Facultad de F\'{i}sica, Pontifica Universidad Cat\'{o}lica de Chile, 306, Santiago 22, Chile}
\altaffiltext{6}{Millennium Institute of Astrophysics, Vicu\~{n}a Mackenna 4860, 7820436 Macul, Santiago, Chile} 
\altaffiltext{7}{Space Science Institute, 4750 Walnut Street, Suite 205, Boulder, CO 80301, USA}

%\vspace{-10mm}

\begin{abstract}
Multi-wavelength analysis of the young massive cluster \ymc\ is presented for the first time. Our \chan\ survey of this region enabled the detection of three X-ray emitting stellar members of the cluster, as well as a possible diffuse X-ray component that extends a few arcseconds from the cluster core with an intrinsic flux of (9$\pm$3)$\times10^{-14}$\,\ergcms\ in the 0.5--10\,keV band. Infrared spectra we obtained for two of these X-ray point sources show absorption lines typical of the atmospheres of massive O stars. The X-ray spectrum from the visible extent of  \ymc\, i.e., a 15$^{\prime\prime}$-radius around the cluster, can be modeled with an absorbed power law with \nh\ $= (6_{-3}^{+4})\times10^{22}$\,\cmsq\ and $\Gamma =$ 2$\pm$1. In addition, the X-ray core of \ymc\ coincides with diffuse emission seen in the infrared band and with a local maximum in the radio continuum map. A possible association with a neighboring H\,II region would place \ymc\ at a distance of around 11\,kpc; on the far side of the Norma Arm. At this distance, the cluster is 0.8\,pc wide with a mass density of (1--4)$\times10^{3}$\,$M_{\odot}$\,pc$^{-3}$.
\end{abstract}

%%__________________________________________________________________


\section{Introduction}

Young Massive Clusters (YMCs) are loose collections of tens to thousands of stars more massive than the Sun \citep[see][for a recent review]{por10}. While they are a common feature of starburst galaxies, the Milky Way contains relatively few YMCs, only a small fraction of which are known X-ray emitters or have been (or can be) resolved into individual objects over several wavebands \citep[e.g.,][and references therein]{tow11a}. The most massive Galactic YMC is Westerlund~1 which is located 4\,kpc away in the direction of the Norma Arm \citep[e.g.,][]{cla05}. Members of its stellar population have been resolved in several wavebands, including in the X-rays (by \chan) where evidence was found for X-ray emission from massive, evolved (i.e., post-main-sequence) stars such as OB supergiants and Wolf-Rayet (WR) stars, from colliding wind binaries, and from an anomalous X-ray pulsar \citep{ski06,mun06a,cla08}. 

X-ray emitting YMCs with resolvable populations represent rare opportunities for observational tests of stellar evolution models including those involving the initial mass function and stellar and cluster dynamics, as well as for understanding the energetic feedback processes between massive stars and their environment. As the more massive stars tend to sink to the center of the cluster, they can pair up into massive binaries. Dynamical interactions between this core population can lead to the tightening of the binary (some of which can evolve into an X-ray binary) or even to its ejection from the cluster \citep[e.g.,][and references therein]{por07,por10}. The distribution of the X-ray emitting point sources compared with the normal stellar content can be used to glean clues to the evolutionary history of the cluster.

We used \chan\ to construct a high-resolution X-ray map of a section of the Norma Arm. Among the new objects discovered in the survey were multiple X-ray sources consistent with a YMC named \ymc\ \citep{bor11} from the Vista Variables in the Via Lactae Survey \citep[VVV:][]{min10}. In this paper, we describe the multi-wavelength properties of \ymc\ by complementing the soft X-ray data with dedicated and archival observations gathered in the radio and infrared bands. The data and analytical methods are described in \S\,\ref{sec_obs}, results are presented in \S\,\ref{sec_res}, and we discuss their implications in \S\,\ref{sec_disc}. A summary of our key findings is given in \S\,\ref{sec_conc}.

\section{Observations \& Data Analysis}
\label{sec_obs}

\subsection{X-ray Data}

In 2011 June, the field of \ymc\ was observed with \chan\ during a survey of the Norma Arm (PI: Tomsick; observation IDs \dataset[ADS/Sa.CXO#obs/12529]{12529} and \dataset[ADS/Sa.CXO#obs/12530]{12530}). An additional observation (\dataset[ADS/Sa.CXO#obs/15625]{ObsID 15625}) was performed in early 2013 as part of \chan\ Director's Discretionary Time in which \ymc\ was serendipitously in the field. Table\,\ref{tab_log} contains the journal of these observations.

\begin{deluxetable*}{ c c c c c }
\tablewidth{0pt}
\tabletypesize{\scriptsize}
\tablecaption{\emph{Chandra} observations}
\tablehead{
\colhead{observation ID} & \colhead{start date}  & \colhead{end date} & \colhead{exposure time (s)} & \colhead{off-axis angle of \ymc} }
\startdata

\dataset[ADS/Sa.CXO#obs/12529]{12529}	& 2011-06-16 06:58:05 & 2011-06-16 12:31:35 	& 19014 	& 3\farcm6	\\

\dataset[ADS/Sa.CXO#obs/12530]{12530} 	& 2011-06-16 12:31:35 & 2011-06-16 18:09:14 & 19260 	& 8\farcm5	\\

\dataset[ADS/Sa.CXO#obs/15625]{15625} 	& 2013-03-23 08:17:49 & 2013-03-23 11:54:43 & 9839 	& 6\farcm3 

\enddata
\label{tab_log}
\end{deluxetable*}

Reprocessing and reduction of the \chan\ data relied on CIAO v.4.5 and HEASoft 6.13. Detailed descriptions of the imaging analysis including background, exposure, and astrometric corrections, as well as the source-detection procedure are presented in \citet{for14}. These cleaned event files and their associated lists of detected sources were used as the inputs for this study. Source coordinates are given in the J2000.0 epoch with uncertainties quoted at 90\% confidence. 

Spectra were extracted from each event file in the 0.5--10\,keV band for source regions (of variable sizes, see \S\,\ref{sec_res}) and from a source-free background region (a rectangle with dimensions: $200^{\prime\prime}\times100^{\prime\prime} = $ 400\,px$\times$200\,px) on the same detector chip. Parameters from absorbed power laws fit to the individual spectra were statistically consistent with each other (i.e., there was no significant X-ray variability between 2011 and 2013). Therefore, the X-ray spectra described in the following sections are based on summing these individual spectra with \texttt{combine\_spectra}.

%-----------------------------Figure Start------------------------------
\begin{figure*}[!t]
\begin{center}
\includegraphics[width=\textwidth]{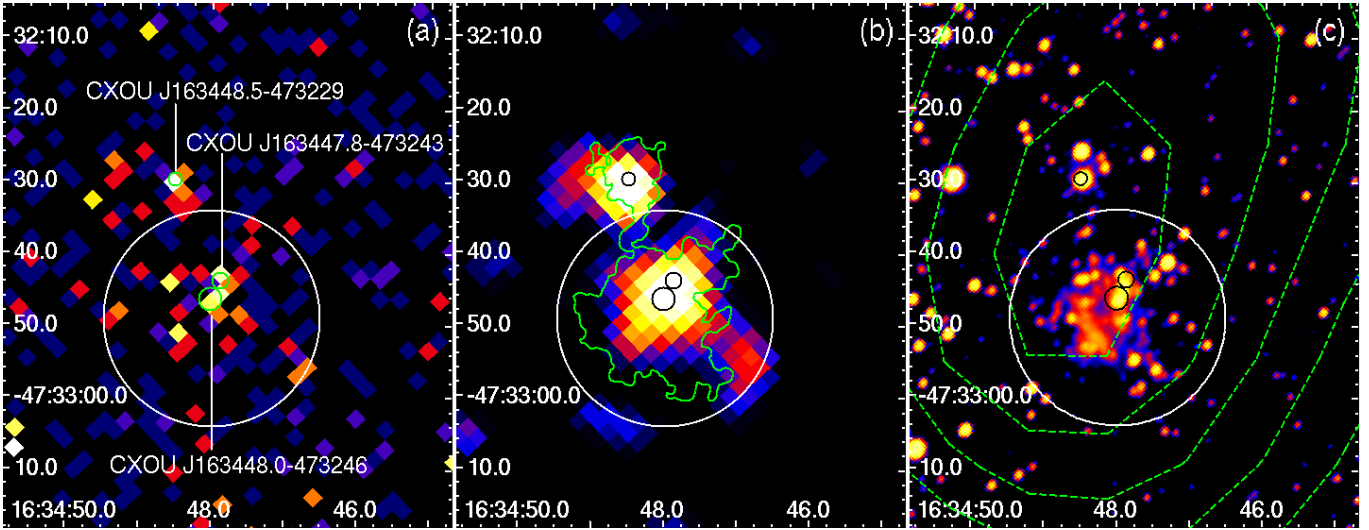}
\end{center}
\caption{\footnotesize{Images of the young massive cluster VVV~CL077 in X-rays (0.5--10\,keV) from \chan-ACIS-I (panels \emph{a} and \emph{b}) and in the infrared ($K_{s}$) band from the VVV Survey \citep[panel \emph{c}:][]{min10}. Each image covers the same region of the sky presented in equatorial (J2000.0) coordinates where North is up and East is left. The X-ray mosaic images combine ObsIDs 12529 and 12530, they are astrometrically corrected, background subtracted, rebinned with $2^{\prime\prime}$-wide pixels, and logarithmically scaled. The X-ray image in panel \emph{b} has been smoothed with a Gaussian ($\sigma=3$\,px) while the infrared image is scaled with a histogram equalization. Detected X-ray sources are shown with their CXOU designations and circle sizes representing the position uncertainty (at 90\% confidence, see Table\,\ref{tab_src}). The large circle (radius $= 15^{\prime\prime}$) represents the visual extent of the YMC according to \citet{bor11}. Contours overlaid on the smoothed X-ray image denote the outline of diffuse infrared emission from the cluster (and bright point sources), while the contours on the $K_{s}$-band image signify radio intensities at 843\,MHz from the MGPS-2 Survey \citep{mur07}. }}
\label{fig_ymc}
\end{figure*}
%-----------------------------Figure End--------------------------------

\subsection{Infrared Data}

The infrared data consist of observations we gathered (PI: Rahoui) at the Cerro Tololo Inter-American Observatory (CTIO). Photometry was performed on 2011 July 19 with the NOAO/NEWFIRM instrument on the 4-m Blanco telescope. Simultaneous $J$, $H$, and $K_{s}$-band spectra (R$\,\,\sim1200$) were collected on 2012 June 1--4 for five stars near the core of the \ymc\ using the Ohio State Infrared Imager/Spectrometer (OSIRIS) mounted on the 4.1-m Southern Astrophysical Research (SOAR) Telescope. 

A set of standard stars was observed in similar conditions in order to remove telluric absorption lines. The ABBA nodding technique was employed for background subtraction. Data reduction relied on the IRAF \textsc{echelle} package which included bad-pixel correction, dark subtraction, linearity correction, flat-fielding, sky subtraction, and spectral extraction along the dispersion axis. The spectra were then wavelength-calibrated by comparing them to an argon lamp. Additional details regarding the observation and analysis procedures will be described in \citet{rah14}. 

Unfortunately, weather and atmospheric conditions were poor with thin clouds and a seeing higher than 1\farcs3. Useful spectra were obtained for only 2 of the 5 targets, and only in the $K$-band. These data are supplemented by archival imaging observations from the VIRCAM instrument of the Vista Variables in the Via Lactae Survey  \citep[VVV:][]{min10}.

\section{Results}
\label{sec_res}

\subsection{X-ray Data}
\label{sec_resx}

In Fig.\,\ref{fig_ymc}, we present images of the field of \ymc\ in the X-rays (panels \emph{a} and \emph{b}: 0.5--10\,keV from \chan) and in the near infrared ($K_{s}$-band image from the VVV Survey \citep{min10}). Detected X-ray sources are indicated by circles with radii equal to the position uncertainty at 90\% confidence (see Table\,\ref{tab_src}). The large circle (radius $=15^{\prime\prime}$) denotes the visual extent of the YMC in the infrared according to \citet{bor11}; it is centered at R.A. $=16^{\mathrm{h}}34^{\mathrm{m}}48^{\mathrm{s}}$ and Decl. $= -47^{\circ}32^{\prime}49^{\prime\prime}$.

This figure illustrates the complex morphology of the multi-wavelength emission from \ymc. In the infrared, the cluster contains 21 visually-identified members \citep{bor11}, plus an extended component from unresolved stars and diffuse emission. In the X-rays, there are 3 detected sources not including extended emission. The size and shape of the extended emission in X-rays and infrared are similar and both have positions that are consistent with a local maximum in the radio continuum map from the Molonglo Galactic Plane Survey 2nd Epoch \citep[MGPS-2:][]{mur07}.

Two candidate X-ray sources are detected near the core of \ymc\ (Table\,\ref{tab_src}). The first source (in R.A.) is \object{CXOU J163447.8$-$473243}, is detected at a significance of $5.8\sigma$ at R.A. $=16^{\mathrm{h}}34^{\mathrm{m}}47^{\mathrm{s}}.88$ and Decl. $= -47^{\circ}32^{\prime}43\farcs8$ with an error radius of 1\farcs1. The other, \object{CXOU J163448.0$-$473246}, is detected at $6.9\sigma$ at R.A.$=16^{\mathrm{h}}34^{\mathrm{m}}48^{\mathrm{s}}.02$ and Decl. $= -47^{\circ}32^{\prime}46\farcs3$ with an error radius of 1\farcs6.

An angular distance of 2\farcs9 separates the two X-ray objects, which is somewhat larger than the 90\% encircled energy radius (i.e., 2\farcs4 at 2.3\,keV) of the Chandra point-spread function (PSF). Their X-ray positions are statistically incompatible at 90\% confidence, and they are consistent with two bright nodes (stars belonging to \ymc) at opposite ends of a clump of unresolved emission in the infrared images. Given the positional coincidence with stars seen in the infrared, we conclude that these are likely to be two distinct X-ray sources rather than a single extended source.

\begin{deluxetable*}{ l c c c c }
\tablewidth{0pt}
\tabletypesize{\scriptsize}
\tablecaption{X-ray sources detected in the vicinity of VVV~CL077}
\tablehead{
\colhead{name} & \colhead{R.A. (deg)}  & \colhead{decl. (deg)} & \colhead{90\% conf. radius} & \colhead{det. sig. ($\sigma$)}}
\startdata

\object{CXOU~J163447.8$-$473243} 	& 248.69950 & $-$47.54550 	& 1\farcs1 & 5.8	\\

\object{CXOU~J163448.0$-$473246} 	& 248.70008 & $-$47.54619 	& 1\farcs6 & 6.9	\\

\object{CXOU~J163448.5$-$473229} 	& 248.70221 & $-$47.54161 	& 0\farcs9 & 6.2

\enddata

\label{tab_src}
\end{deluxetable*}

To confirm whether any of the X-ray emission emanating from the \ymc\ is extended or simply arises from two point sources, we used ray-tracing simulation data from ChaRT (part of the CIAO package) to generate reprojected PSFs for the observation in which the source had the smallest off-axis angle (ObsID \dataset[ADS/Sa.CXO#obs/12529]{12529}). For this simulation, we adopted an exposure time of 19\,ks, and an absorbed power law spectrum with $\Gamma=2.2$ and \nh\ fixed at $6\times10^{22}$\,\cmsq\ (see below, and in Table\,\ref{tab_spec}). The CIAO tool {\tt srcextent} returns an intrinsic source size of (7\farcs0 $\pm$ 2\farcs5) for this emission, assuming the object is placed on-axis. Using an observation with a larger off-axis angle (ObsID \dataset[ADS/Sa.CXO#obs/12530]{12530}) leads to a larger intrinsic source size ($\sim 15^{\prime\prime}$), as expected.

After subtracting off 95\% of the photons from the two X-ray point sources in the core via aperture restriction (see below), we find that there is still some residual emission with as many net counts outside the core (44) as from the two point sources combined (40). Along with the large intrinsic size of the X-ray emitting region, this indicates the presence of an extended and diffuse X-ray component, in addition to the emission from the two point sources.

Just outside the YMC is \object{CXOU J163448.5$-$473229} which coincides spatially with a star that is likely associated with the YMC given the similar infrared colors to those of other cluster members. The best-fitting X-ray position ($6.2\sigma$ detection significance) is R.A. $=16^{\mathrm{h}}34^{\mathrm{m}}48^{\mathrm{s}}.53$ and Decl. $= -47^{\circ}32^{\prime}29\farcs8$ with an error radius of 0\farcs9. correc{Its intrinsic size, estimated via ChaRT, is consistent with 0$^{\prime\prime}$ in ObsID \dataset[ADS/Sa.CXO#obs/12529]{12529}.}

Note that in ObsID \dataset[ADS/Sa.CXO#obs/12530]{12530}, a second candidate X-ray source is detected outside the YMC, \object{CXOU J163448.6$-$473234}, located at R.A. $=16^{\mathrm{h}}34^{\mathrm{m}}48^{\mathrm{s}}.62$ and Decl. $= -47^{\circ}32^{\prime}34\farcs1$ with an error radius of 2\farcs5 (significance $=5.1\sigma$). The large estimate of the source extent from \texttt{wavdetect} ($=10^{\prime\prime}$), the fact that it was only detected in this ObsID when the YMC was 8\farcm5 off axis, and the lack of a clear infrared counterpart suggest that this is the same source as \object{CXOU J163448.5$-$473229} (4\farcs4 away), and so it will not be discussed further.

Spectral data were fit with power law, \texttt{apec}, and bremsstrahlung models. However, the observed photon counts within the source extraction regions are low (i.e., $\lesssim100$\,counts), and so we caution that any derived spectral parameters are merely estimates, and may not be indicative of the true values. Given the low source counts and very low background counts, spectral fitting relied on \citet{cas79} statistics and \citet{pea00} $\chi^{2}$ test statistics. All fits assumed \citet{wil00} abundances and photoionization cross sections from \citet{bal92}. The hardness ratio is defined as $(H-S)/(H+S)$ where $H$ is the count rate in 3--10\,keV and $S$ is the count rate in 0.5--3\,keV. 

\begin{figure}[!t]
\begin{center}
\includegraphics[width=0.3\textwidth,angle=-90]{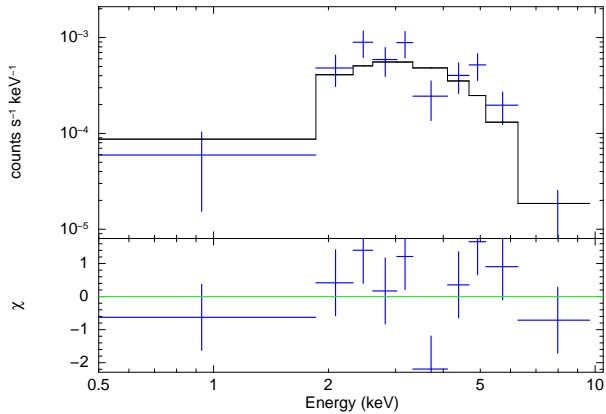}
\end{center}
\caption{X-ray spectrum extracted from a $15^{\prime\prime}$ radius (30 pixels) around the infrared position of \ymc. The lower panel shows residuals from fitting an absorbed power law (solid curve) to the spectral data. For visual clarity, each spectral bin contains a minimum of 15 counts. Spectral parameters from the unbinned analysis are listed in Table\,\ref{tab_spec}. }
\label{fig_xspec}
\end{figure}

The X-ray spectrum extracted from a $15^{\prime\prime}$ radius (30 pixels) around the infrared position of \ymc\ is presented in Fig.\,\ref{fig_xspec}. This spectrum includes the two X-ray point sources in the core of the YMC, plus diffuse X-ray emission. There are 107 net counts in this region for a hardness ratio of 0.10$\pm$0.11. A power law fit to the data yields \nh\ $=(6.4_{-2.9}^{+3.7})\times10^{22}$\,\cmsq\ and $\Gamma = 2.2_{-1.0}^{+1.2}$, while an apec model yields \nh\ $=(5.6_{-2.1}^{+3.2})\times10^{22}$\,\cmsq\ and $kT = 4_{-2}^{+8}$\,keV.

To compare the properties of the X-ray emission in the core of the YMC with that of the diffuse emission, we extracted a spectrum from a circle of $6^{\prime\prime}$ radius (12\,pixels) that encompasses the 95\% encircled energy radius (EER) of both point sources in the core (\object{CXOU J163447.8$-$473243} and \object{CXOU J163448.0$-$473246}) giving 40 net source counts, as well as a spectrum from an annulus (centered at the same position) with inner and outer radii of $6^{\prime\prime}$ and 12\farcs5, respectively, that represents the diffuse emission (44 net counts). The results of models fit to these spectra are tabulated in Table\,\ref{tab_spec}. There are hints that the diffuse component has a slightly higher plasma temperature (with apec models) and a harder photon index (with power law models) than the stellar sources in the core region, but the low photon counts results in large uncertainties, which leads to some overlap in the range of possible temperatures and photon indices at the 90\% confidence level. We attempted to extract spectra for each X-ray point source in the core by using non-overlapping regions but the sources ended up with too few counts (9 net counts each) to permit reliable fitting.

%-----------------------------Figure Start------------------------------
%-----------------------------Figure End--------------------------------

\begin{figure}[!t]
\begin{center}
\includegraphics[width=0.5\textwidth]{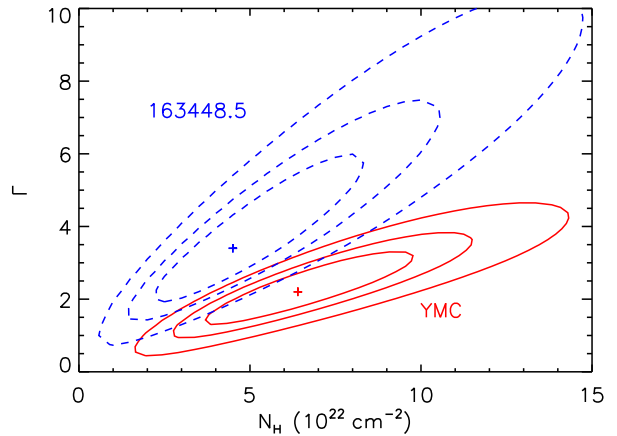}
\end{center}
\caption{Parameter space of \nh\ and $\Gamma$ resulting from absorbed power-law fits to the X-ray spectra (0.5--10\,keV) of selected extraction regions. Two extraction regions are considered: a 15$^{\prime\prime}$-radius circle centered on the IR position of \ymc\ (solid lines), and a 6$^{\prime\prime}$-radius circle centered on the X-ray position of CXOU J163448.5$-$473229 (dashed lines). The best-fitting parameters are represented by crosses and the contours denote confidence levels of 68\%, 90\%, and 99\%.}
\label{fig_con}
\end{figure}

A source spectrum was made for \object{CXOU J163448.5$-$473229} by using an extraction radius of $6^{\prime\prime}$ (12 pixels) centered on its X-ray position. While the spectral parameters of \object{CXOU J163448.5$-$473229} are compatible with those of the other source regions (within the statistical uncertainties), it is clearly softer than the others given its hardness ratio of $-$0.43$\pm$0.25. This is illustrated in Fig.\,\ref{fig_con} which compares the 68\%, 90\%, and 99\% confidence contours of spectral parameters from the power law model fit to the data of \object{CXOU J163448.5$-$473229} (dashed lines), and for the YMC (solid lines). There is some overlap between the parameter spaces at the 99\% confidence level, but the X-ray emission from the cluster tends to be harder and more absorbed than that of a point source situated in the cluster's outskirts.

\subsection{Infrared Data}

\begin{figure}[!t]
\includegraphics[width=0.45\textwidth]{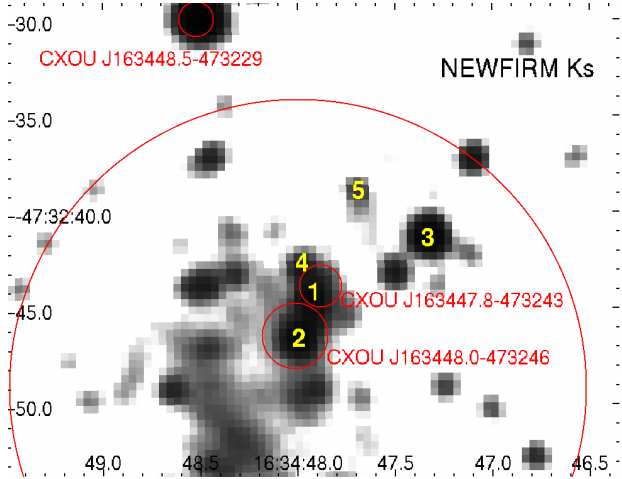}
\caption{\footnotesize{Image of \ymc\ in the infrared $K_{s}$-band from NOAO/NEWFIRM. Circles denote the positional uncertainty of the X-ray point sources described in this work (Table\,\ref{tab_src}), and the visual extent of the YMC according to \citet{bor11}. Infrared objects targeted for follow-up spectroscopy are numbered 1--5.}}
\label{fig_ymc_zoom}
\end{figure}

\begin{figure}[!t]
\begin{center}
\includegraphics[width=0.5\textwidth]{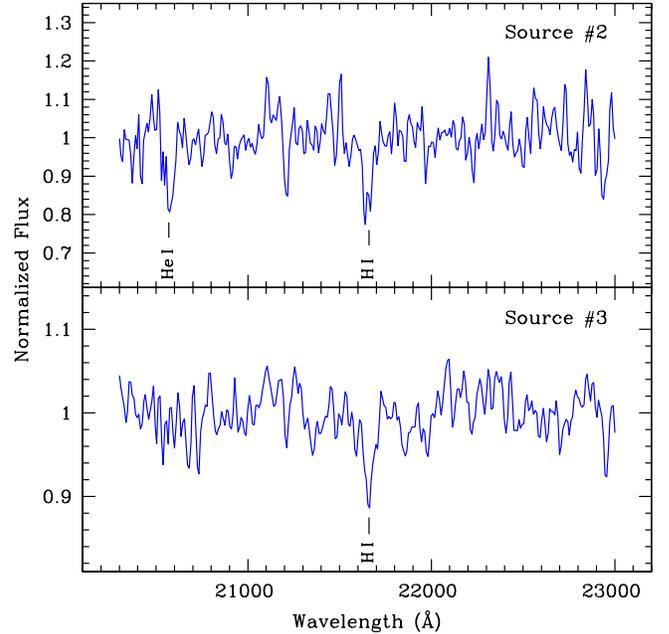}
\end{center}
\caption{Infrared $K_{s}$-band spectra of two stars (labeled ``2'' and ``3'' in Fig.\,\ref{fig_ymc_zoom}) in the YMC VVV~CL077 from observations gathered with NOAO/OSIRIS. }
\label{fig_irspec}
\end{figure}

Figure\,\ref{fig_ymc_zoom} shows the $K_{s}$-band image of the region gathered with NOAO/NEWFIRM. Infrared spectra for Source \#2 and Source \#3 are presented in Fig.\,\ref{fig_irspec}. In the $K_{s}$-band spectra of Source \#2 , we detect a Br-$\gamma$ line in absorption at 21660\,$\AA$. The equivalent width is $11.5$\,$\AA$ with a typical error of $\sim$1--1.5\,$\AA$. There is a marginal detection of the He\,I line in absorption at 20600\,$\AA$. Source \#3 also presents a Br-$\gamma$ line in absorption with an equivalent width of $6.5$\,$\AA$. With standard deviations being less than 0.06, this leads to $S/N > 10$ in both sources. Comparing these spectra to those of a sample of high-mass stars provided by \citet{han96} indicates that Sources \#2 and \#3 are probably late-O giants.

In addition to these spectra, we searched the ViZier archives for multi-wavelength counterparts consistent with the X-ray error circles of the sources detected by \chan. Table\,\ref{tab_ir} lists the reported magnitudes of these counterpart candidates. The derived $J - H$ and $H - K_{s}$ colors are compatible with the colors observed in O stars \citep[e.g.,][]{ram12}. Comparing \object{CXOU J163448.0$-$473246} and \object{CXOU J163448.5$-$473229}, we find consistent values (within the uncertainties) for their $J - H$ (1.8$\pm$0.2 vs. $2.0_{-0.2}^{+0.1}$) and $J - K_{s}$ ($3.3_{-0.2}^{+0.1}$ vs. $3.0_{-0.2}^{+0.1}$) colors, but the $H - K_{s}$ color is $\sim$50\% higher for \object{CXOU J163448.0$-$473246} (1.5$\pm$0.2 vs. $1.0_{-0.1}^{+0.2}$) which suggests it has a slight near-to-mid-infrared excess, although the source is blended in the 2MASS $K_{s}$-band image. We point out the presence of a mid-infrared source located 2\farcs9 away from \object{CXOU J163448.0$-$473246} (i.e., 1\farcs4 from the IR position of the cluster): \object{WISE J163448.13$-$473248.9} has reported magnitudes of 5.563$\pm$0.110 (at 3.35\,$\mu$m), 4.756$\pm$0.055 (4.6\,$\mu$m), -0.896$\pm$0.121 (11.6\,$\mu$m), and -3.190$\pm$0.020 (22.1\,$\mu$m).

%-----------------------------Figure Start------------------------------
%-----------------------------Figure End--------------------------------

\section{Discussion}
\label{sec_disc}

At least 21 stars belonging to \ymc\ were visually identified in the infrared \citep{bor11}, and this young massive cluster is now detected for the first time in the X-rays with \chan. A complete census of the X-ray emitting sources in \ymc\ can not be achieved with the current data. Nevertheless, recent X-ray observations of this region reveal a number of intriguing results. 

Imaging data from \chan\ suggest that there are three X-ray point sources associated with infrared stellar counterparts that belong to the cluster. Two of these sources are located near the cluster core, and the third is located in the cluster's outskirts. In addition to these point sources, there is extended and diffuse X-ray emission: the estimated intrinsic (on-axis) size of the X-ray emitting core of the YMC is larger than expected for the combined 95\% energy-containment circles of the two closely-spaced point sources in the core; and this component, which contains as many net counts as the core X-ray point sources combined, can not be directly associated with any resolved sources.

The presence of diffuse emission, with X-ray spectral properties statistically consistent with that of the resolved stellar X-ray sources, would suggest a population of unresolved X-ray sources in the YMC whose emission lies below the sensitivity limit of the survey \citep[$\sim10^{-15}$\,\ergcms][]{for14}. 

Diffuse X-ray emission has been observed in other YMCs including in Wd~1 and in active star-forming regions like the Carina Nebula \citep{mun06b,tow11b,tow11c}. In addition to the combined emission from unresolved stars, the diffuse component could originate from putative collisions between the winds of massive stars, although the temperature and absorption in \ymc\ are higher than expected for such processes \citep[e.g.,][]{ste03}.

Two of the stars \ymc\ display infrared spectral signatures typical of late-type O stars \citep{han96}. There is good overall agreement between the outlines of the emission as seen in the X-rays and infrared, and this multi-wavelength consistency continues into the radio band where the X-ray and infrared emission coincide spatially with a local maximum in the 843\,MHz continuum map from MGPS-2 \citep{mur07}. The coincidence of the multi-wavelength emission from \ymc\ could indicate a common physical origin, however, the observed X-ray flux of the diffuse component is larger than expected from an extrapolation of the infrared flux density (see below). 

%-----------------------------Figure Start------------------------------
%-----------------------------Figure End--------------------------------

The nearest catalogued ultra-compact H\,II region is \object{IRAS~16311$-$4726} \citep{ira88,bro96} which is 6$^{\prime\prime}$ from the infrared position of \ymc. It has a velocity of 121\,km\,s$^{-1}$ with respect to the local standard of rest, which translates to a source distance of approximately 11\,kpc \citep[e.g.,][]{bra93}. There is also a catalogued H\,II region 7$^{\prime}$ away from \ymc\ with a listed distance of 10.9$\pm$0.2\,kpc \citep{rus03}. If these regions are related, then we can assign a distance of around 11\,kpc to \ymc. This would make it one of the most distant Galactic YMCs known, comparable to \object{Mercer~81} \citep{dav12}, at a location consistent with the far Norma Arm. 

Taking the visual extent of the bulk of the stellar population in the infrared as the angular scale \citep[15$^{\prime\prime}$:][]{bor11} gives a cluster diameter of 0.8\,pc for an assumed distance of 11\,kpc. This value is consistent with the intrinsic sizes of other Galactic YMCs \citep[][and references therein]{fig08}.

Corrected for absorption, the diffuse emission from \ymc\ has an intrinsic flux of $(9\pm3)\times10^{-14}$\,\ergcms\ (0.5--10-keV). This is more than an order of magnitude above the flux expected from the Galactic Ridge diffuse emission \citep{ebi01,ebi05}. If we assume, nevertheless, that the Ridge emission accounts for 20\% of the flux from \ymc, as it does for Wd~1 \citep{mun06b}, then the residual flux of $\sim 7\times10^{-14}$\,\ergcms\ can be attributed to the diffuse cluster emission. For a distance of 11\,kpc (assumed), this translates to an X-ray luminosity of $10^{34}$\,\ergs. Flux densities reported for \object{IRAS~16311$-$4726} are 20.1$\pm$0.6\,Jy at 12\,$\mu$m and 153$\pm$5\,Jy at 25\,$\mu$m \citep{ira88}. Extrapolating a power law with a photon index of $\Gamma = 2.8\pm0.1$ leads to an X-ray flux density of $1.3\times10^{-10}$\,Jy at 1\,keV ($3.0\times10^{-15}$\,\ergcms\ over 0.5--10\,keV). This is an order of magnitude smaller than the lower 90\%-confidence boundary of the intrinsic flux density ($3\times10^{-9}$\,Jy at 1\,keV). In order to match the X-ray flux, the photon index should be no steeper than 2.5, which is consistent with what we found in Section\,\ref{sec_resx}.

%-----------------------------Figure Start------------------------------
%-----------------------------Figure End--------------------------------

A simple estimation of the cluster mass density yields $\rho \sim$(1--4)$\times10^{3}$\,$M_{\odot}$\,pc$^{-3}$. This assumes that all 21 visually-identified members of the cluster are similar to the sources we identified as O type stars with masses in the range of 10--50\,$M_{\odot}$. While the mass density is consistent with those of other YMCs, the cluster mass implied by this first-order approximation ($\sim10^{3}$\,$M_{\odot}$) is an order of magnitude below the inferred masses of other YMCs \citep{fig08}. This indicates a large population of unresolved stars hiding in and around the cluster which is not surprising given the extinction in this direction. 

From the observed infrared excess $E(H - K_{s}) = 1.5\pm0.2$, we estimate the extinction in the optical band as $A_{V} = 15\pm2$ assuming a near-infrared extinction law with $\alpha = -2.1$ \citep{ste09}. Applying the relation of \citet{guv09} yields a column density of \nh\ $=(3.3_{-1.0}^{+1.1})\times10^{22}$\,\cmsq. When the contribution from molecular hydrogen is included, the total column density of the interstellar medium towards the Norma region ranges from $4.9\times10^{22}$\,\cmsq\ to $9.4\times10^{22}$\,\cmsq\ \citep{rah14}. This is consistent with our X-ray-derived value of $(6.4_{-2.9}^{+3.7})\times10^{22}$\,\cmsq.

Plasma temperatures of around 1--2\,keV are found for the X-ray emitting point sources in the cluster which is higher than average for massive stars \citep[e.g.,][]{rau14}. The X-ray source located in the outskirts of the cluster (\object{CXOU J163448.5$-$473229}) has a temperature that is statistically compatible (at 90\% confidence) with 0.7\,keV which is a more typical value. As we move towards the center of the cluster, we find that the X-ray derived temperatures of these stars are higher, possibly due to increasing column density (which preferentially absorbs low-temperature photons). With its proximity to \object{IRAS~16311$-$4726}, this could indicate that the stars in the core of \ymc\ are still embedded within their dusty natal cocoons; similar to the hot, young stars of W49A \citep{con02}, a cluster that resides within an ultra-compact H\,II region. However, we can not exclude that the unusual spectral values are the result of poor photon statistics.

Based on the IR spectra, we narrowed the range of spectral types to O5--O9 giants for two of the stars in the cluster. We derived the distance modulus $m - M = 5 \log(d) + A - 5$ using the observed $J$-band magnitude of \object{CXOU J163448.0$-$473246} in Table\,\ref{tab_ir}, the absolute $J$-band magnitude of $-$4.6 for an O giant \citep{mar06}, and $A_{J} = 0.28 A_{V}$. This gives a distance to the object of 10$\pm$3\,kpc, consistent with the distance assumed based on its possible link with \object{IRAS~16311$-$4726}. 

In massive stars, the ratio between the X-ray and bolometric luminosities is $\sim 10^{-7}$ \citep[e.g.,][]{ber97,rau14}. Such stars have bolometric luminosities of $\log(L/L_{\odot}) \sim$ 5.2--5.7 \citep{mar05}. To test whether the objects we classify as massive stars obey this rule, we began with the observed X-ray flux of the ``inner core'' region and corrected this for the interstellar absorption alone (\nh\ $= 3\times10^{22}$\,\cmsq), then halved this value to obtain the absorption-corrected flux of one of the stars of the pair. For an assumed distance of 10$\pm$3\,kpc, this yields an X-ray luminosity of $3.5\times10^{32}$\,\ergs. A ratio $L_{\mathrm{X}}/L_{\mathrm{bol}}$ of $10^{-7}$ is obtained for either the high end of the bolometric luminosity range ($\log(L/L_{\odot}) \sim$5.7), or for distances at the lower end of the estimate ($\sim$7\,kpc). Fixing the temperature to a more reasonable value of 0.6\,keV raises the column density to unlikely values (\nh\ $=(22_{-5}^{+7})\times10^{22}$\,\cmsq), which, once corrected for interstellar absorption, results in  $L_{\mathrm{X}} \sim 2.5\times10^{32}$\,\ergs, and so it remains consistent (within the uncertainties) with the expected $L_{\mathrm{X}}/L_{\mathrm{bol}}$ value for massive stars.

Finally, we consider the question of whether there are supernova remnants and accreting compact objects in \ymc. There are no obvious shell-like features in the infrared and X-ray data that would indicate recent supernova activity, although the low level of soft X-ray emission and lack of hard X-ray or gamma-ray detections appear to rule out such scenarios. Luminous ($L_{X}\gtrsim10^{38}$\,\ergs) X-ray point sources are missing from \ymc\ which excludes high-mass X-ray binaries (HMXBs) where the accretion proceeds through Roche-lobe overflow. Wind-fed HMXBs could be present but they would have to be weakly-accreting or quiescent ($L_{X}\sim10^{33}$\,\ergs), such as is the case for Supergiant Fast X-ray Transients (SFXTs) outside of outbursts. However, the photon indices measured in the core and for the external X-ray sources are too soft to support the weakly-accreting SFXT scenario \citep[e.g.,][]{rom11}. An additional point to consider is that the supernova that leads to the formation of the compact object in such systems can impart a velocity to the nascent X-ray binary that is several times larger than the cluster escape velocity \citep{van00,cla05}. The apparent lack of supernova remnants or accreting compact objects would tend to favor a relatively young age for the YMC, i.e., a few Myr at most.

\section{Summary \& Conclusions}
\label{sec_conc}

We presented results from recent \chan\ observations of \ymc, a Young Massive Cluster (YMC) located towards the Norma Arm. Three X-ray point sources were detected that correspond to visually-identified stellar members of the cluster. Infrared spectra we obtained for two members of the cluster show absorption lines typical of the atmospheres of massive stars. Plus, the infrared colors and $L_{\mathrm{X}}/L_{\mathrm{bol}}$ relation also indicate massive stars. However, X-ray spectral fits of these stars yield temperatures that are higher than average for massive stars.

In addition to these point sources, there are hints that an extended region of diffuse X-ray emission permeates the cluster core. The X-ray column density and optical extinction, as well as an estimate of the cluster mass, suggests a population of unresolved stars that contribute to the diffuse emission. The X-ray spectrum of the 15$^{\prime\prime}$-radius visible extent of the \ymc\ is best modeled with an absorbed [\nh\ $= (3-10)\times10^{22}$\,\cmsq] power law ($\Gamma = 2\pm1$). 

The X-ray core of the cluster coincides with diffuse emission seen in the infrared band and with a local maximum in the radio continuum map. A tentative link with a neighboring H\,II region would place \ymc\ at a distance of around 11\,kpc; on the far side of the Norma Arm. At this distance, the cluster is 0.8\,pc wide with a mass density of (1--4)$\times10^{3}$\,$M_{\odot}$\,pc$^{-3}$, i.e., within the range of values seen for other YMCs. 

A complete picture of the X-ray population and diffuse emission from \ymc\ is not possible at this time, but continued X-ray observations of YMCs are important for the study of massive stars, the formation and evolution of compact objects, and the recent star formation history of the Galaxy.

\acknowledgments
The authors thank the anonymous referee whose careful, patient, and constructive review led to significant improvements in the manuscript. This work was supported in part through \chan\ Award Number GO1-12068A issued by the \chan\ X-ray Observatory Center, which is operated by the Smithsonian Astrophysical Observatory for and on behalf of the National Aeronautics Space Administration under contract NAS8-03060. FEB acknowledges support from Basal-CATA (PFB-06/2007) and CONICYT-Chile (under grant FONDECYT 1141218 and Anillo ACT1101). The scientific results reported in this article are based on observations made by the \chan\ X-ray Observatory. This research has made use of: software provided by the \chan\ X-ray Center (CXC) in the application packages CIAO, ChIPS, and Sherpa; data obtained from the High Energy Astrophysics Science Archive Research Center (HEASARC) provided by NASA's Goddard Space Flight Center; NASA's Astrophysics Data System Bibliographic Services; the SIMBAD database operated at CDS, Strasbourg, France; data products from the Two Micron All Sky Survey, which is a joint project of the University of Massachusetts and the Infrared Processing and Analysis Center/California Institute of Technology, funded by NASA and the NSF; data products from the Wide-field Infrared Survey Explorer, which is a joint project of the University of California, Los Angeles, and the Jet Propulsion Laboratory/California Institute of Technology, funded by NASA and the NSF.
%.

{\it Facilities:} \facility{CXO}, \facility{Blanco (NEWFIRM)}, \facility{SOAR (OSIRIS)}

\bibliographystyle{apj}
\bibliography{bod.bib}

\clearpage

\begin{deluxetable}{ l c c c c c c c c c c  }
\tablewidth{0pt}
\tabletypesize{\scriptsize}
\tablecaption{Parameters from absorbed power law and apec models fit to the unbinned X-ray spectra of selected regions}
\tablehead{
  & \colhead{\nh\ \tablenotemark{a}}  & \colhead{$\Gamma$\tablenotemark{b}} & \colhead{$kT$\tablenotemark{c}} & \colhead{norm.\tablenotemark{d}} & \colhead{C-stat./goodness\tablenotemark{e}} & \colhead{$S$\tablenotemark{f}} & \colhead{$H$\tablenotemark{g}} & \colhead{HR\tablenotemark{h}} & \colhead{obs. flux\tablenotemark{i}} & \colhead{unabs. flux\tablenotemark{j}}  }
\startdata	

\noalign{\smallskip}
\sidehead{\ymc: 15$^{\prime\prime}$ extraction radius centered on IR position of \citet{bor11}}

\smallskip
			& $6.4_{-2.9}^{+3.7}$ & $2.2_{-1.0}^{+1.2}$ & ---	& 5.4 	& 73\% & 48$\pm$6 & 59$\pm$10 & 0.10$\pm$0.11 & $6.8_{-1.8}^{+2.4}$ & $21.6_{-5.9}^{+10.9}$ \\

\smallskip\smallskip

			& $5.6_{-2.1}^{+3.2}$ & ---			 & $4_{-2}^{+8}$	& 9.9 & 78\% & --- & --- & --- & $6.6_{-1.7}^{+2.3}$ & $15.0_{-4.5}^{+16.1}$	\\

\noalign{\smallskip}
\sidehead{\ymc\ diffuse emission: annulus with 6$^{\prime\prime}$--12\farcs5 radii centered between CXOU J163447.8$-$473243 and CXOU J163448.0$-$473246 }

\smallskip
			& $5.4_{-3.5}^{+5.0}$ & $1.7_{-1.3}^{+1.6}$ & ---	& 1.4 	& 73\% & 16$\pm$5 & 28$\pm$7 & 0.27$\pm$0.20 & $4.6_{-1.7}^{+2.8}$ & $8.9_{-2.5}^{+3.1}$ \\

\smallskip\smallskip

			& $5.9_{-2.4}^{+3.7}$ & ---			 & $5_{-3}^{+26}$	& 5.4 & 82\% & --- & --- & --- & $4.1_{-1.3}^{+1.7}$ & $8.9_{-3.2}^{+8.8}$	\\

\noalign{\smallskip}
\sidehead{\ymc\ inner core: circle with 6$^{\prime\prime}$ radius centered between CXOU J163447.8$-$473243 and CXOU J163448.0$-$473246}

\smallskip
			& $8.0_{-4.6}^{+8.9}$ & $3.1_{-1.6}^{+2.4}$ & ---	& 9.2 	& 61\% & 18$\pm$5 & 22$\pm$5 & 0.10$\pm$0.18 & $2.8_{-0.9}^{+1.5}$ & $27.7_{-6.9}^{+8.1}$ \\

\smallskip\smallskip

			& $9.0_{-6.3}^{+8.5}$ & ---			 & $1.6_{-0.4}^{+1.0}$	& 16.8 & 70\% & --- & --- & --- & $2.4_{-0.7}^{+1.3}$ & $19.6_{-14.8}^{+21.1}$	\\

\noalign{\smallskip}
\sidehead{\object{CXOU J163448.5$-$473229}: circle with 6$^{\prime\prime}$ radius}

\smallskip
			& $4.5_{-2.5}^{+4.3}$ & $3.4_{-1.6}^{+2.9}$ & ---	& 4.5 	& 3\% & 20$\pm$5 & 8$\pm$4 & $-$0.43$\pm$0.25 & $1.3_{-0.5}^{+0.9}$ & $13.1_{-6.9}^{+22.3}$ \\

\smallskip\smallskip

			& $4.9_{-2.0}^{+2.2}$ & ---			 & $1.2_{-0.5}^{+1.2}$	& 8.2 & 2\% & --- & --- & --- & $1.1_{-0.4}^{+0.6}$ & $12.0_{-8.6}^{+36.6}$	\\

\enddata
\tablenotetext{a}{Column density in units of $10^{22}$\,\cmsq. }
\tablenotetext{b}{Photon index of the power law model.}
\tablenotetext{c}{Plasma temperature (keV) for the apec model.}
\tablenotetext{d}{Model normalization ($\times10^{-5}$). }
\tablenotetext{e}{Goodness of fit defined as the percentage of simulated spectra (of 10,000 trials) which return a statistic smaller than the observed value.}
\tablenotetext{f}{Net source counts in the soft ($S$) band: 0.5--3\,keV.}
\tablenotetext{g}{Net source counts in the hard ($H$) band: 3--10\,keV.}
\tablenotetext{h}{Hardness ratio defined as $(H-S)/(H+S)$.}
\tablenotetext{i}{Observed flux (i.e., not corrected for absorption) in units of $10^{-14}$\,\ergcms\ in the 0.5--10\,keV band.}
\tablenotetext{j}{Absorption-corrected flux in units of $10^{-14}$\,\ergcms\ in the 0.5--10\,keV band.}

\label{tab_spec}
\end{deluxetable}

\clearpage
\begin{turnpage}

\begin{deluxetable}{ l l c c c c c c c c c }
\tabletypesize{\scriptsize}
\tablecaption{Infrared counterpart candidates to X-ray sources near VVV~CL077}
\tablehead{
\colhead{name} & \colhead{counterpart} & \colhead{offset ($^{\prime\prime}$)} & \colhead{$J$}  & \colhead{$H$} & \colhead{$K_{s}$} & \colhead{3.6\,$\mu$m}  & \colhead{4.5\,$\mu$m} & \colhead{5.8\,$\mu$m}  }
\startdata

\noalign{\smallskip}

\object{CXOU~J163447.8$-$473243} 			& ---\tablenotemark{$\dagger$} & --- & --- & --- & --- & --- & --- & --- 	\\

\noalign{\smallskip}

\object{CXOU~J163448.0$-$473246} 			& \object{2MASS~J16344796$-$4732452} & 1.2 & 14.485$\pm$0.087 
										& 12.700$\pm$0.137 & 11.230$\pm$0.081 & --- & --- & ---   	\\
										& \object{DENIS~J163447.9$-$473245} 	& 0.8 & 14.10$\pm$0.11 
										& --- & 11.12$\pm$0.07 & --- & --- & ---   		\\
										& \object{VVV~515726976188} & 0.2 & 14.731$\pm$0.005 & 13.466$\pm$0.005 
										& --- & --- & --- & ---   		\\
					
\noalign{\smallskip}
				
\object{CXOU~J163448.5$-$473229}			& \object{2MASS~J16344850$-$4732297} & 0.3 & 13.729$\pm$0.075 
										& 11.772$\pm$0.075 & 10.730$\pm$0.043 & --- & --- & ---   	\\
										& \object{VVV~515726978416} & 0.6 & 13.647$\pm$0.002 & 11.898$\pm$0.001 
										& --- & --- & --- & ---   		\\
										& \object{GLIMPSE~G336.8914$+$00.0524} & 0.4 & --- & --- & --- 
										& 10.052$\pm$0.048 & 9.844$\pm$0.053 & 9.818$\pm$0.103

\enddata

\tablecomments{Infrared magnitudes of counterpart candidates in: the $J$, $H$, and $K_{s}$ bands from 2MASS \citep{skr06}, DENIS \citep {den05}, and VVV \citep{min10}; and at 3.6\,$\mu$m, 4.5\,$\mu$m, and 5.8\,$\mu$m from \emph{Spitzer}-GLIMPSE \citep{gli09}.}

\tablenotetext{$\dagger$}{The nearest IR source with catalogued magnitudes, \object{2MASS~J16344796$-$4732452}, is $1\farcs7$ away and is likely associated with \object{CXOU~J163448.0$-$473246}.}

\label{tab_ir}
\end{deluxetable}

\end{turnpage}

\clearpage
\global\pdfpageattr\expandafter{\the\pdfpageattr/Rotate 90}

\end{document}